\documentclass[prl,twocolumn,superscriptaddress,showpacs]{revtex4}

\usepackage{bm,amsmath,amssymb,graphicx}

\begin{document}

\title{ The curvature of the rotating disk and its quantum manifestation}

\author{Victor Atanasov}
\affiliation{Faculty of Physics, Sofia University, 5 blvd J. Bourchier, Sofia 1164, Bulgaria}
\email{vatanaso@phys.uni-sofia.bg}
\homepage[web page ]{http://phys.uni-sofia.bg/~vatanaso/}

\author{Rossen Dandoloff}
\affiliation{ Laboratoire de Physique Th\'{e}orique et
Mod\'{e}lisation , Universit\'{e} de Cergy-Pontoise, F-95302 Cergy-Pontoise, France}

\begin{abstract}
The geometry of the rotating disk is revisited and the quantum consequences are discussed. A suggestion to detect the presence of the Gaussian curvature on the rotating disk only measuring transition frequencies is made. A quantum equivalent of the Newtonian bucket is considered.
\end{abstract}

\pacs{03.30.+p, 03.65.-w, 04.20.-q, 74.50.+r}

\maketitle

In this paper we propose a method to prove the existence of curvature in a rotating 2D disc. The consideration of the metrics of a rotating disk has led Einstein\cite{Einst}
to believe that the geometry of the rotating disk is non-Euclidean. At the origin of the discussion about the rotating disk is the so called Ehrenfest paradox\cite{Ehren}. 
Ehrenfest has suggested that because of the Lorentz contraction of the rim of the disk and of course of all the radii from the rim inwards ( with different degree
of contraction depending on the linear velocity), the disk will bent and acquire positive Gaussian curvature. This point of view later was contradicted by Einstein. He claimed that we should not consider the process of acceleration of the disk, but rather the disk which already rotates with steady angular velocity $\omega$.
Here we follow the discussion of this problem by D.Dieks \cite{Dieks}. The simplest approach to this problem is based on the consideration of the line element. In the Minkowski space it is given by:
$$ds^2 = c^2dt^2 - dx^2 - dy^2 - dz^2 = c^2dt^2 - dr^2 - r^2d\phi^2 - dz^2$$
It is only natural to use cylindrical coordinates which are stationary in the rotating disk. The coordinate transformation  from the stationary to the rotating coordinate system is: t=t', r=r', z=z' and $\phi = \phi' +\omega t'$, where the primed coordinates are in the rotating coordinate system. The rotation is performed around
the z=z' axis. In the rotating frame the line element is given by:
$$ ds^2 = (c^2 - r'^2\omega^2)dt'^2 - dr'^2 - r'^2d\phi'^2 - dz'^2 - 2\omega r'^2d\phi'dt' $$

In what follows we will drop the prime on the coordinates.
Now, we will use the fact  that a light signal follows a null-geodesic i.e. $ds=0$. From this equation we can calculate the time a light signal travels from $A$ to $B$ and back. In the
usual Minkowski space this gives $dt_{1,2}=\pm1/c\sqrt{dr^2+r^2d\phi^2+dz^2}$ i.e. the light signal travels from $A$ to $B$ for the time interval $dt_1$ and it travels
from $B$ to $A$ for the time $dt_2$ (where now $c<0$ because the signal travels in the oposite direction). If $t_1$ corresponds to $A$
and $t_2$ corresponds to $B$, (where $t_1-t_2=dt$) the time difference $t_1-t_2$ is a measure of the simultaneity at $A$ and $B$.
In the non-rotating Minkowski space $dt=t_1-t_2=1/c\sqrt{dr^2+r^2d\phi^2+dz^2}-1/c\sqrt{dr^2+r^2d\phi^2+dz^2}=0$. If we put $dt=0$
in the expression for the Minkowski line element $ds^2$ we get the line element in the corresponding 3-dimensional space
$dl^2=dr^2+r^2d\phi^2+dz^2$.

 In the rotating coordinate system the equation $ds=0$ gives the two intervals for the travel from $A$ to $B$ and from $B$ to $A$ as follows:
$$dt_1= \frac{\omega r^2d\phi + \sqrt{(c^2-\omega^2r^2)(dz^2+dr^2)+c^2r^2d\phi^2}}{(c^2-\omega^2r^2)} $$
and 
$$dt_2= \frac{-\omega r^2d\phi + \sqrt{(c^2-\omega^2r^2)(dz^2+dr^2)+c^2r^2d\phi^2}}{(c^2-\omega^2r^2)} $$

Following Dieks simultaneity is defined using only light signals between $A$ and $B$. If $t_0$ is the time coordinate of the emission of the light signal at $A$, then simultaneous events at $A$ and $B$ have the following time coordinates: $t_0+1/2(dt_1+dt_2)$ as measured at $A$ and $t_0+dt_1$ as measured at $B$.

It follows then that the time difference between events that are measured by synchronized clocks is given by the following expression:
$$dt=(t_0+dt_1) - (t_0 +1/2(dt_1 +dt_2))=$$
$$1/2(dt_1-dt_2)= \frac{\omega r^2d\phi}{c^2-\omega^2r^2}$$

Note that if $d\phi=0$ the two events lie on the radius and $dt=0$ because there are no relativistic effects along the radius of the rotating disk.
Now, the spacial distance between two infinitesimally close points (as measured in the rotating disk) is found by substituting the time difference as calculated above (which represents the measure for simultaneity) into the expression for the line element $ds$. The result is the following line element on the rotating disk:
$$ds^2 = dr^2 + \frac{r^2d\phi^2}{1-\omega^2r^2/c^2} =dr^2 + h(r)^2d\phi^2$$
Here $ds$ is the line element in the 2D rotating disc and $h(r)=\frac{r}{1-\omega^2r^2/c^2}$  is the corresponding Lam\'e coefficient. It is obvious that the geometry of the rotating disc is non Euclidean. The corresponding Gaussian curvature is given by:
$$K = - \frac{1}{h}\frac{\partial^2h}{\partial r^2} $$
Usually $\frac{\omega}{c}<<1$, then in the leading term in $\frac{\omega}{c}$ the Gaussian curvature 
$$K= - \frac{3\omega^2}{c^2}.$$
It is negative and approximately constant. Negative Gaussian curvature is the equivalent of the effect of negative mass affecting the curvature of space-time.

The ratio $\frac{\omega}{c}$ being very small, the detection of this Gaussian curvature appears to be very delicate task. The detection of this Gaussian curvature is
important because the controversy about the rotating disk continues even today.

Now we will turn our attention to the quantum manifestation of the rotating disk. The influence of the rotation of the disk in the Schrödinger equation is two-fold: first the Gaussian curvature modifies the potential and second the rotation modifies the linear momentum via Galilean transformation. We will start by considering the Schrödinger equation on a surface with a given Gaussian curvature.

We note that the polar coordinate system represents an example of ${\it half-geodesic\, coordinates}$ \cite{geodesic}. Curves with constant radii correspond to concentric circles with centers at $r=0$. These are called {\it geodesically parallel} \cite{geodesic} to the circle which represents the boundary of the removed disk. (we may remove a small disk around the origin of the coordinate system in order to avoid problems at the origin of the coordinate system where the polar coordinates are not defined). The radial coordinate traces out geodesics. The line-element of such a coordinate system where one of the coordinate lines corresponds to geodesics is in general given by $ds^2=dr^2 + h(r)^2d\phi^2$ where $\phi$ is the geodesic coordinate. The function $h (r )$ is a function only of this coordinate.  Let us briefly consider this point in some generality. In these half-geodesic coordinates the two-dimensional Laplacian $\Delta_2$ has the following form
\begin{equation}\label{eq.1}
 \Delta_2 \Psi = \frac{1}{h}\left[\frac{\partial }{\partial r}\left(h\frac{\partial h}{\partial r}\right) + \frac{\partial}{\partial \phi}\left(\frac{1}{h}\frac{\partial h}{\partial \phi}\right)\right] \Psi \, .
\end{equation}
The normalization integral of the wave-function is given by $\int|\Psi|^2\sqrt{g}dr d\phi$. Here $g$ is the determinant of the metric tensor.

Let us turn to the stationary Schr\"odinger equation in the half-geodesic coordinate system. In the case of  polar coordinates the Schr\"odinger equation will exhibit a term which contains the first derivative of the wave-function with respect to the geodesic coordinate.  In order to derive an equation without such a term we choose to factorize the wave function in the following way $\Psi=g^{-1/4}\psi$. In this expression $g$ is the determinant of the metric tensor such that $g=h^2$. The Schr\"odinger equation is then turned into the following form
\begin{equation}
-\frac{\hbar^2}{2M}\Delta_2\left(\frac{\psi}{\sqrt h}\right)=E\frac{\psi}{\sqrt h}\Rightarrow
\end{equation}
$$
-\frac{\hbar^2}{2M}\psi'' + \frac{\hbar^2}{2M}\frac{1}{4h}\frac{\partial^2 h}{\partial r^2}\psi - \frac{\hbar^2}{2M}\frac{1}{2h^2}\left(\frac{\partial h}{\partial r}\right)^2\psi
$$
\begin{equation}
 + \frac{\hbar^2}{2Mh^2}\frac{\partial^2 \psi}{\partial\phi^2}=E\psi\, . 
\end{equation}
Here $(\psi'')$ stands for $\frac{\partial^2\psi}{\partial r^2}$. The above  equation may be written as follows:
\begin{equation}
-\frac{\hbar^2}{2M}\psi'' + V(r,\phi)\psi=E\psi
\end{equation}
Curves with $r$=constant represent isometric directions. Since these curves are also closed, the wave-function $\psi$ must be a periodic function in $\phi$ and we may simply write $\psi(r,\phi)=\psi_1(r)e^{i{\rm m}\phi}$ where ${\rm m}$ represents the quantized angular momentum. 
In the effective potential $V(r,\phi)$ we recognize the intrinsic Gaussian curvature of the surface $K=-h^{-1}\partial^2_r h$ and the geodesic curvature of the geodesically parallel coordinate lines $k_g=-h^{-1}\partial_r h$. Now the effective potential has the form\cite{RD}:
\begin{equation}
V=\frac{\hbar^2}{2M}\bigl[\frac{{\rm m}^2}{h^2}-\frac{1}{4}k_g^2 - \frac{1}{2}K\bigr]
\end{equation}
In the limit $\frac{\omega}{c}<<1$, $k_g^2=\frac{1}{r^2} + 2\frac{\omega^2}{c^2}$ and $K = -3\frac{\omega^2}{c^2}.$

In this limit the effective potential on a surface with the above Gaussian curvature has the following form:
\begin{equation}\label{QEP}
V = \frac{\hbar^2}{2M}\bigl[\frac{{\rm m}^2-1/4}{r^2}-({\rm m}^2 - 1)\frac{\omega^2}{c^2}\bigr]
\end{equation}

Now let us consider the full problem of a rotating disk. Here we will use the above result  about the effective potential due to the Gaussian curvature but now we will take into account the fact that the linear momentum is transformed under rotation with a constant angular velocity $\vec\omega$ using a Galilean transformation which leads to the following expression for the kinetic energy:

\begin{equation}
\frac{1}{2M} \left[ \vec{p} -  M \vec{\omega} \times \vec{r}      \right]^2 
\end{equation}

The quantum version of this new linear momentum is obtained in the usual way by replacing $\vec{p}$ with $ \frac{\hbar}{i} \vec{\nabla}$, where $\vec{\nabla}$ is given in the $(r,\psi,z)$ coordinates.
Note that $\vec{\omega}=const$ and $\vec{r}$ are multiplicative operators. Then we obtain the following Hamiltonian for a free particle on the rotating disk:

\begin{equation}
\nonumber H=\frac{1}{2M} \left[ \frac{\hbar}{i} \nabla -  M \vec{\omega} \times \vec{r}      \right]^2 =
\end{equation}

$$
-\frac{\hbar^2}{2M} \Delta + i \frac{\hbar}{2} \left[ {\rm div}(\vec{\omega} \times \vec{r} ) + (\vec{\omega} \times \vec{r}) . \; \nabla     \right] + \frac12 M (\vec{\omega} \times \vec{r})^2
$$

In the cylindrical coordinates we are using:   $\vec{\omega}=(0, 0, \omega)$ and $\vec{r}=(r, 0, 0).$ Then we obviously have:
\begin{equation}
\nonumber \vec{\omega} \times \vec{r}= (0, r \omega, 0) \quad {\rm div}(\vec{\omega} \times \vec{r} ) = 0
\end{equation}

The Hamiltonian now reads:

\begin{equation}
H=-\frac{\hbar^2}{2M} \Delta + i \frac{\hbar}{2} r \omega \frac{1}{h(r)} \frac{\partial }{\partial \phi}   + \frac12 M r^2 \omega^2
\end{equation}
and with $ h(r)=\frac{r}{1-\omega^2r^2/c^2}$ this expression becomes:
\begin{equation}
H=-\frac{\hbar^2}{2M} \Delta + i \frac{\hbar}{2}  \omega \left( 1-\omega^2r^2/c^2 \right) \frac{\partial }{\partial \phi}   + \frac12 M r^2 \omega^2
\end{equation}
Here $\Delta$ has the same form as in eq. (\ref{eq.1}).
Now we renormalize the wave function and use the rotational symmetry of the problem. We make the following ansatz for the wave function: 
$$\Psi(r,\phi)=\frac{\psi(r) }{\sqrt{(r)} }e^{i {\rm m}\phi}$$

This leads to the following radial Schrodinger equation:
$$
-\frac{\hbar^2}{2M} \psi''+ V(r)\psi=E\psi 
$$
Where $V(r)$ represents the effective potential:
\begin{eqnarray}\label{q.bucket}
V(r) &=&\frac{\hbar^2}{2M}\left[\frac{{\rm m}^2-1/4}{r^2}-({\rm m}^2 - 1)\frac{\omega^2}{c^2}\right] \\
\nonumber &&- \,  \hbar \omega \frac{{\rm m} }{2}   \left( 1-r^2\frac{\omega^2}{c^2} \right)  + \frac12 M r^2 \omega^2 
\end{eqnarray}
Note that here the rotation lifts the degeneracy in $m$ namely for $\omega=0$ \, $V_m(r)=V_{-m}(r)$. For the experimental verification of the Gaussian curvature we  propose to explore the dependency of the  potential difference $\Delta V= V_{m+1} - V_{m}$ on $\omega:$ 
\begin{eqnarray}
\Delta V & =& \left[   \frac{\hbar^2}{2 M} \frac{2 {\rm m}  +1 }{r^2}  - \hbar \omega  \right] \left( 1-r^2\frac{\omega^2}{c^2} \right) 
\end{eqnarray}
Here the transition energy between the levels is linear in $\omega$ and the relativistic correction is contained in the   factor $ \left( 1-r^2\frac{\omega^2}{c^2} \right)$ which stemms from the Gaussian curvature of the metric manifested in the effective potential. The measurement of the relativistic correction to $\Delta V$ will give us unequivocal answer to the question of the presence of Gaussian curvature on the rotating disk.

In conclusion we would like to comment on the quantum equivalent of the Newtonian bucket gedanken experiment, namely the water placed in a rotating bucket gradually is being draged into rotation with an angular velocity colinear with that of the bucket. Expression (\ref{q.bucket}) suggests a similar behaviour on the quantum level, namely the energy of the $-{\rm |m|}$ state is higher than that of the ${\rm |m|}$ state provided $\omega$ is positive. Note that ${\rm m}>0$ corresponds to a rotation of the quantum particle in the same directtion as the rotating disk, i.e. the rotating disk ''drags'' along the quantum particles  as in the classical case (because these states are more energetically favourable than ${\rm m}<0$, which correspond to a rotation in the opposite direction; if $\omega=0$ both ${\rm m}$ states correspond to the same energy and rotations in both directions are equally probable).

\end{document}